# Google's Training Supercomputers from TPU v2 to Ironwood: Architectural Stability, Scale, Resilience, Power Efficiency, and Sustainability Across Five Generations

Norman P. Jouppi, Sridhar Lakshmanamurthy, Cliff Young, and David Patterson, Google, LLC

## Abstract

This paper (to appear in the July/August 2026 issue of *IEEE Micro* magazine) summarizes five generations of Google's TPUs, from TPU v2 to Ironwood, highlighting their evolution as scalable, resilient, power-efficient, sustainable supercomputers for AI training. It details the TPU's stable architecture, which has surprisingly easily accommodated the rapidly changing deep neural network workloads, such as the rise of Transformers. Key advancements over eight years include 10x increase in HBM capacity and bandwidth per node, a 100x increase in peak node performance, and a 3600x increase in supercomputer performance. The paper also discusses the role of optical circuit switches, built-in self test, and hardware replay in enhancing resilience and how TPU's environmental impact is reduced with substantial improvements in performance per Watt and in carbon emissions per floating point operation. It concludes by identifying six features that may well characterize successful training accelerators of this decade.

## Introduction

A decade ago, few expected a software company like Google to build its own chips. The announcement in May 2016 of the first *Tensor Processing Unit* (*TPU*) startled many. It offered 30x the performance per Watt of contemporary GPUs and 80x over contemporary CPUs for deep neural network (DNN) inference [Jouppi17, Jouppi18]. TPU v1 inspired others to act:

- Within 4 months Intel spent billions of dollars acquiring two DNN hardware startups (Nervana and Movidius) and then two more over the next few years (MobilEye and Habana).
- Over 18 months the venture community invested $1.5B in 45 startups [Metz18], growing to ~$3B annually in over 100 DNN hardware startups.
- Hyperscalers Alibaba and Amazon (and eventually Microsoft) started their own DNN inference chips.

Its widespread influence motivated one pundit to observe that TPU v1 "launched a thousand chips" [Babbge24], a tongue-in-cheek reference to the famous line about Helen of Troy.[1]

TPU v2, Google's first training supercomputer, was already well underway in May 2016. Twelve months later it was deployed and revealed publicly. Since then, Google has continuously deployed new chips for DNN training and serving, powering key services.

This paper reviews five generations of TPU training supercomputers[2], also called *pods*. Skeptics initially warned that an ASIC might be too tailored to existing DNN models, quickly becoming outdated given AI's rapid pace. That proved not to be the case. The founding principles of TPU v2 demonstrated remarkable longevity, with later generations increasing component speed and size by riding technological breakthroughs without altering the underlying design. Not all accelerators can make this claim.

In addition to documenting TPU's staying power, this paper illustrates how it scaled successfully. TPUs have hurdled the Accelerator Wall [Fuchs19] that claims Moore's Law accounts for the majority of benefit after the first couple of generations. In an era of diminished Moore's Law and no Dennard scaling, TPU supercomputers grew a staggering 3600x in system-level performance. The paper also explains how their resilience improved to work effectively at 36x more nodes, how their carbon footprint per floating point operation shrank, and pinpoints features that may explain TPU's durable success.

---

[1] "Was this the face that launch'd a thousand ships," from *Doctor Faustus*, by Christopher Marlowe, 1594.
[2] Trillium (TPU v6) is focussed on inference.

## Improved Scale Over Time

Table 1 shows how the TPU node and supercomputer scaled over 8 years:

- HBM (High Bandwidth Memory) capacity and bandwidth per TPU increased ~10X;
- Performance per Watt rose ~30X;
- Supercomputer size and interconnect bisection bandwidth both grew ~40X;
- Directly-addressable shared (supercomputer-level) HBM memory expanded ~400x, from 4 terabytes in TPU v2 to 1.77 petabytes in Ironwood, a new record for AI supercomputers;
- Peak performance per TPU advanced a remarkable ~100X, from 46 BF16 TFLOPS in TPU v2 to 4614 FP8 TFLOPS in Ironwood (or ~50X for BF16 at 2307 TFLOPS);
- Peak supercomputer performance jumped ~3600X, an impressive compound annual growth rate of nearly 100% despite losing Dennard scaling and Moore's Law.

| | TPU v2 | TPU v3 | TPU v4 | TPU v5p | Ironwood |
|---|---|---|---|---|---|
| Year | 2017 | 2018 | 2021 | 2023 | 2025 |
| Peak TPU BF16 TFLOPS | 46 | 123 | 275 | 459 | 2307 |
| Peak TPU BF8 TFLOPS | N.A. | N.A. | N.A. | 459 | 4614 |
| Matrix Multiply Units (MXU) per TPU (Number, Size) | 2 128x128 bf16 | 4 128x128 bf16 | 8 128x128 bf16 | 8 128x128 bf16 | 4 256x256 bf16, 4 512x512 fp8 |
| VMEM per TPU (MiB) | 32 | 32 | 32 | 128 | 128 |
| HBM version | HBM2 | HBM2 | HBM2 | HBM2E | HBM3E |
| Number HBM stacks per TPU | 2 | 4 | 4 | 6 | 8 |
| HBM capacity per TPU (GiB) | 16 | 32 | 32 | 96 | 192 |
| HBM BW (GB/sec) | 700 | 900 | 1200 | 2765 | 7300 |
| TensorCores per TPU | 2 | 2 | 2 | 2 | 2 |
| SparseCores per TPU | 2 | 2 | 4 | 4 | 4 |
| Cooling | Air | Liquid | Liquid | Liquid | Liquid |
| TPUs per CPU host | 4 | 8 | 4 | 4 | 4 |
| Pod size (Number of TPUs) | 256 | 1024 | 4096 | 8960 | 9216 |
| Pod Topology | 2D Torus | 2D Torus | 3D Torus | 3D Torus | 3D Torus |
| ICI links per Node | 4 | 4 | 6 | 6 | 6 |
| ICI BW per Link (GB/sec) | 62 | 70 | 50 | 100 | 100 |
| Pod Bisection BW (GB/sec) | 1984 | 4480 | 25,600 | 64,000 | 76,800 |
| Pod peak BF16 ExaFLOPS | 0.01 | 0.13 | 1.1 | 4.1 | 21.3 |
| Pod peak BF8 ExaFLOPS | N.A. | N.A. | N.A. | 4.1 | 42.5 |
| Pod HBM Capacity (PetaBytes) | 4 | 33 | 131 | 851 | 1769 |
| Relative Pod TFLOPS (Normalized FP8 FLOPS) | 1 | 10 | 100 | 350 | 3600 |
| Relative Pod TFLOPS/W (TDP) | 1 | 1.8 | 4.9 | 5.2 | 29.3 |
| Relative Pod TDP | 1 | 5.6 | 20 | 67 | 123 |

**Table 1. TPU node and supercomputer ("pod") features over five generations of training TPUs.**

## Architecture Stability Over Time

Figure 1 shows that DNNs have changed drastically and quickly. Today, variations of Transformer dominate Google workloads. Despite the potential downside of domain specific architectures becoming mismatched to latest DNN trends over the 2–3 years it takes to design, fabricate, and deploy an accelerator, the original TPU v2 microarchitecture demonstrated long-term viability in this fast moving field. Architectural stability reduces the difficulty of optimizing software and models for new TPUs.

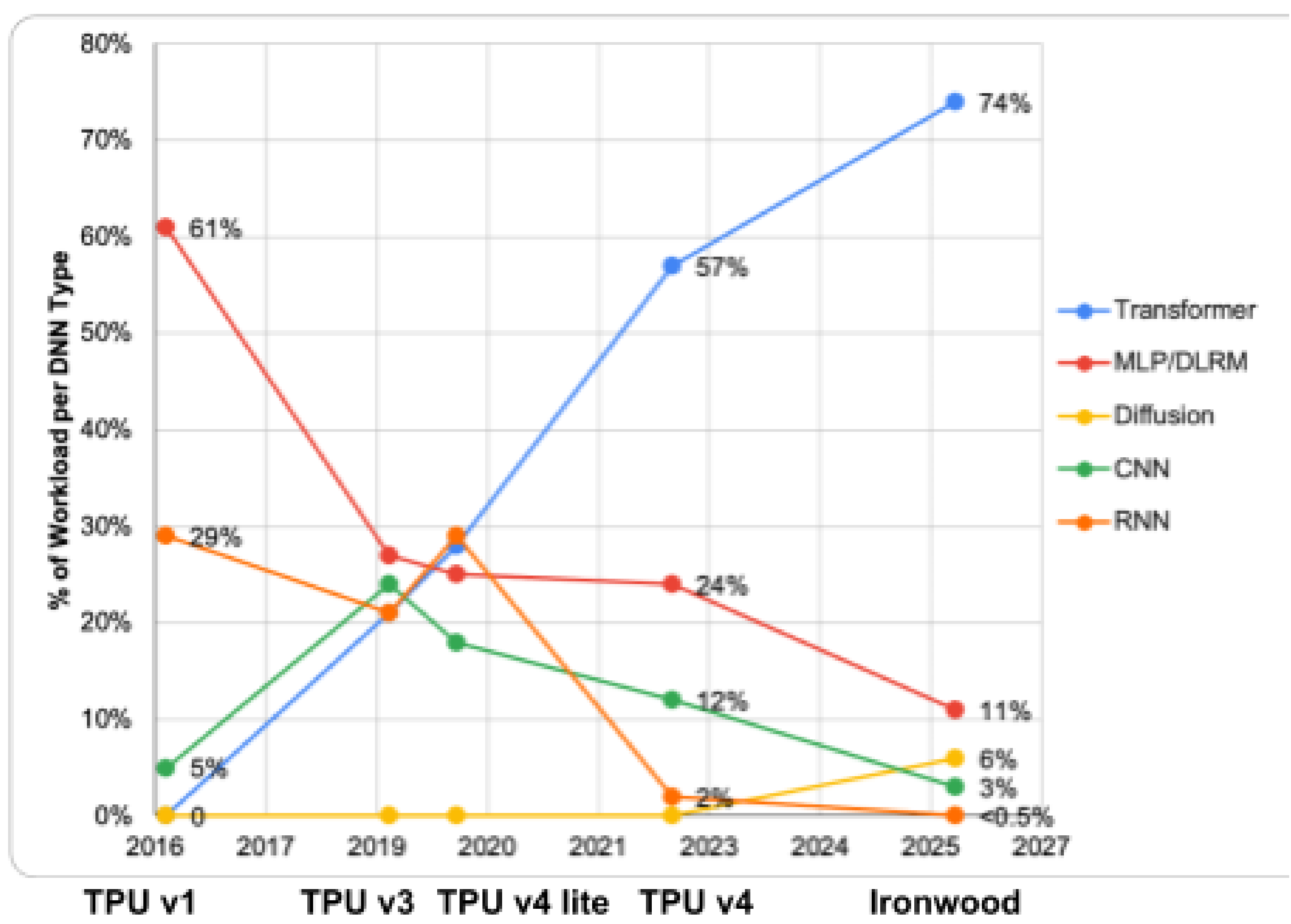

**Figure 1. DNN workload mix over five generations of TPUs [Jouppi23].** The rightmost points are Google's internal workload for training TPUs in 2026. (TPU v1 and TPU v4 lite are TPUs for inference.) In 2016 DNNs were dominated by MLPs at 61%, while in 2019-2020 DNN types were relatively equally balanced. Since 2016, RNNs have practically disappeared (<0.5%), and Diffusion models are now larger than CNNs. The Transformer paper wasn't published till December 2017, yet just 15 months later it was already 21% of Google's production workload. In 2026 DNN workloads are imbalanced again, with the dominant DNN being Transformer at 74%.

Figure 2 shows a TPU v2 block diagram [Norrie21] that remarkably still holds for every training TPU through Ironwood. We first review TPU v2 before describing its evolution. (The next few paragraphs are derived from [Norrie21].)

TPU v2 has two **TensorCores**. Two cores struck the right balance between longer wire latencies of a single large core versus exposing numerous tiny cores that software must lash together. Operating on large blocks of data via two big cores is a much simpler programming model.

TensorCore's **scalar unit** fetches complete ***VLIW (Very Long Instruction Word)*** bundles of 322 bits from a local instruction memory, executes the scalar operation slots locally and then forwards decoded instructions to the vector and matrix units for later execution, decoupled from scalar execution. After scalar execution, the instruction bundle and scalar register values are forwarded to the **vector unit**, which features 128 vector lanes. Each lane includes an additional 8-way data parallel dimension, known as a *sublane*, enabling operations on 8 sets of 128-wide vectors per clock cycle. Each lane's register files perform loads and stores against its local slice of vector memory (**VMEM**). An asynchronous **DMA (Direct Memory Access) unit** transfers data between HBM and local vector memory. The memory hierarchy is compiler-controlled, unlike some other accelerators that rely on caches. Since HBM holds vectors and matrices, DMAs can stride through memory. Upon DMA completion, a notification lands in the core's sync flags, allowing the program to stall until data arrives.

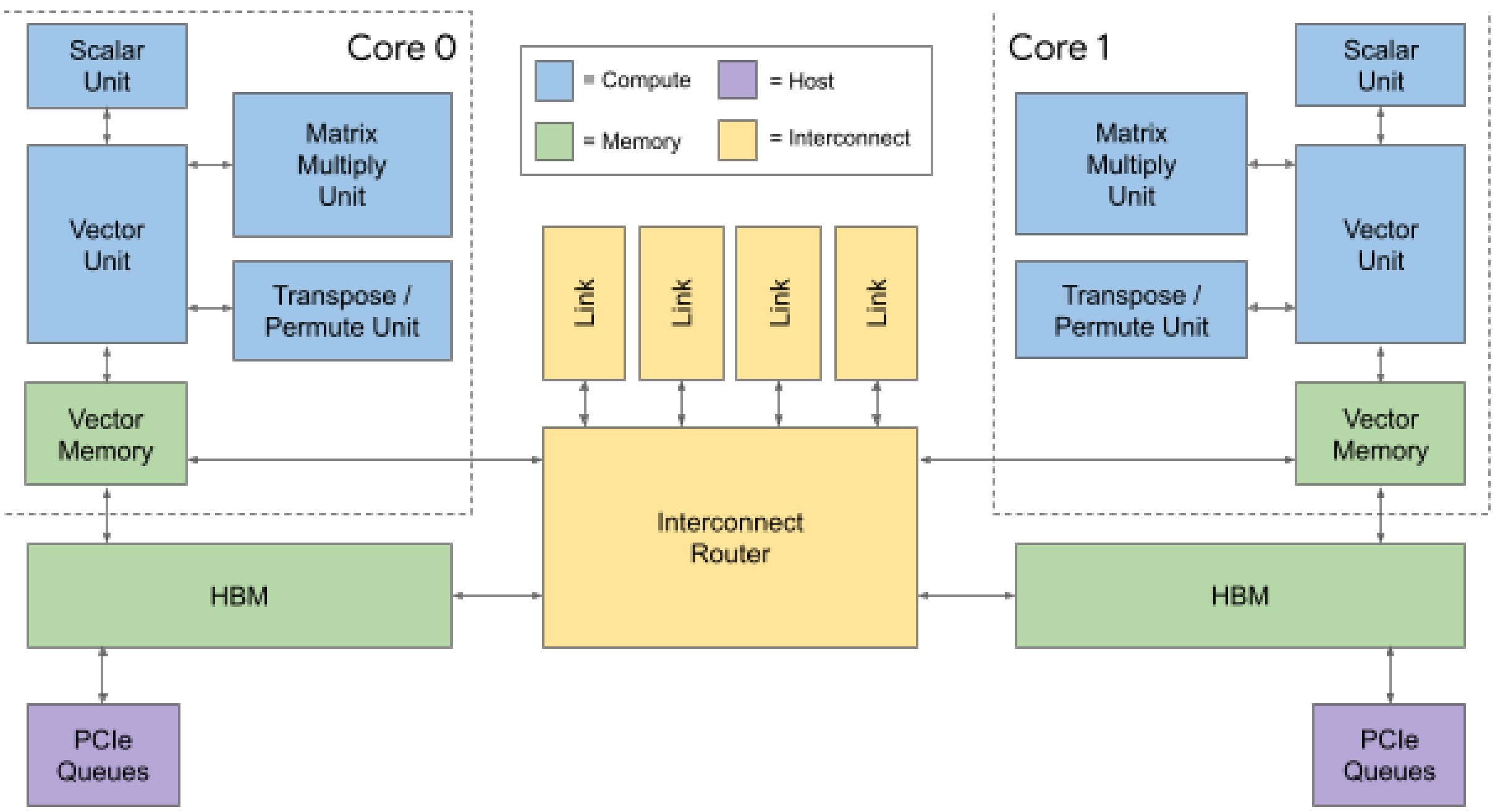

**Figure 2: Block diagram of the two TensorCores in TPU v2 [Norrie21].** The TPU v2 core datapath is blue, HBM is green, host connectivity is purple, and the interconnect router and links are yellow. Later TPUs have more interconnect links and matrix multiply units. The SparseCore was present in TPU v2 but was not disclosed until the TPU v4 paper [Jouppi23]; if it were in this diagram, it would have had links similar to the two TensorCores.

The **matrix multiply unit (MXU)** is the computational heart of TPUs. In TPU v2 it was a 128x128 *systolic array* of multipliers and adders, delivering 32,768 operations per cycle. TPU v2 was the first DNN accelerator to diverge from the IEEE floating point standard, as Google determined that for DNNs range was more important than precision. In 16-bit *Brain Float format (BF16)*, for the first time the exponent (8 bits) is *larger* than its fraction (7 bits). In contrast, IEEE's exponent is smaller, e.g., FP16 uses 5-bit exponents and 10-bit fractions and FP32 uses 8-bit exponents and 23-bit fractions. Many subsequent narrower floating formats have followed BF16's precedent of bigger exponents. Multiplications are in BF16 with accumulation in full IEEE FP32. Beyond matrix multiplication, other units efficiently perform various matrix primitives, such as transposes, row reductions, or column permutations.

TPU v2 featured **four off-chip links** (***Inter-Chip Interconnect*** or ***ICI***) and two on-chip links to an on-chip router. These four links enable the 2D torus system interconnect for 256 TPUs, supporting common ML communication patterns, like AllReduce. DMAs to other TPUs function similarly to DMAs to local HBM, though with a "push-only" restriction for simplicity.[3] This dedicated TPU interconnect enabled scalable synchronous training across all TPUs.

[3] Push-only means it supports writes but not reads. TPUs rely on software to ensure safety and synchronization.

The **SparseCore** is a domain-specific architecture initially for embedding training [Jouppi23]. SparseCores are relatively inexpensive, typically occupying only ~5% of the die area and ~5% of the power. They operate in a sea-of-cores configuration, integrating supercomputer-scale HBM and ICI to create a flat, globally addressable memory space. Unlike AllReduces of large parameter tensors in dense training, the all-to-all transfers of smaller embedding vectors utilize HBM and ICI with finer-grained access patterns for scatter/gather. As separate cores, SparseCores enable parallelization across dense compute, SparseCore, and ICI communications. We consider SparseCore as a "dataflow" architecture because data flows from memory to various specialized compute units.

SparseCore units contain 16 compute tiles. Each tile features an associated HBM channel and supports multiple outstanding memory accesses. Each tile also includes a *Fetch Unit*, a programmable 8-wide SIMD *Vector Processing Unit*, and a *Flush Unit*. The Fetch Unit reads activations and parameters from the HBM into the tile's slice of a 2.5 MiB *Sparse Vector Memory*. During the backward pass, the Flush Unit writes updated parameters to HBM. Similar to TPU v1, the units execute CISC-like instructions and operate on variable-length inputs, where instruction execution time is data-dependent.

SparseCore was initially developed for sparse embedding tables within deep learning recommendation models, used in advertising, search ranking, YouTube, and Google Play applications. These models constituted 61% of the TPU v1 workload in 2016 and still represented ~25% for TPU v4 in 2022 [Jouppi23]. With the rise of Transformer models, which accounted for 60% of the workload by 2022, SparseCores also began serving as offload engines for collective operations like AllReduce, AllGather, ReduceScatter, and Broadcast; data summarization operations like Top-K; and small sparse tensor operations like Transformer Decode. Furthermore, SparseCores enhance performance by operating in parallel while the TensorCores handle dense attention on the feed forward path of Transformers.

We connect the TPU supercomputer to storage over the data center network to provide input data for the model via a PCIe-connected CPU host. Maintaining system balance across CPU, network, and storage is critical for achieving end-to-end performance at scale.

## TPU Software Stack

The TPU software stack has evolved over TPUs from v2 to Ironwood, with the XLA (Accelerated Linear Algebra) framework as the one constant. Early TPUs were driven by TensorFlow, with a "bridge" that translated from TensorFlow graphs into XLA's High Level Optimizer (HLO) format. Inside of XLA, "fusions" (akin to region formation in instruction level parallelism compilers) allow optimization across many operations to save passes through memory. Today JAX (Just-in-time Auto-differentiated XLA) has become the language and system of choice for programming TPUs, with the Pallas kernel language adding fine-grained control for model developers.

## Architecture Evolution Over Time

Figure 3 shows five generations of TPU boards and packages for the three most recent. It illustrates the remarkable stability of always having four TPUs per board despite tremendous changes. The first three images capture the jump from air to liquid cooling and the switch from a liquid cooling loop to distribution and collection manifolds. Less visible are the advances in power delivery and regulation (including vertical power delivery) or the increase in die size, chiplet count, and packaging sophistication.

The evolutions of TPU microarchitecture across generations is primarily in scale and number of components rather than in introducing new microarchitecture features, as in other accelerators:

- **TensorCores:** Every training TPU uses two physical TensorCores that share only HBM. Since TPU v4, our XLA (Accelerated Linear Algebra) compiler has supported tensor parallelization directives that give the illusion of a single large core—called *Megacore*—unifying HBM capacity and ICI bandwidth in a single effective thread [JAX24].
- **Matrix multiply unit (MXU):** Systolic arrays proved to be a foundational building block that most accelerators either adopted from the start or eventually integrated if they didn't. Leveraging improved logic density from 16 nm to finer geometries, MXUs scaled from two 128x128 systolic arrays in TPU v2 to four 256x256 arrays for bf16 in Ironwood. Ironwood also added support for FP8 arithmetic, which means it can also compute four 512x512 FP8 multiplies. Similar to adding redundant rows to improve memory yield and cost, Ironwood added a redundant row in the MXU.

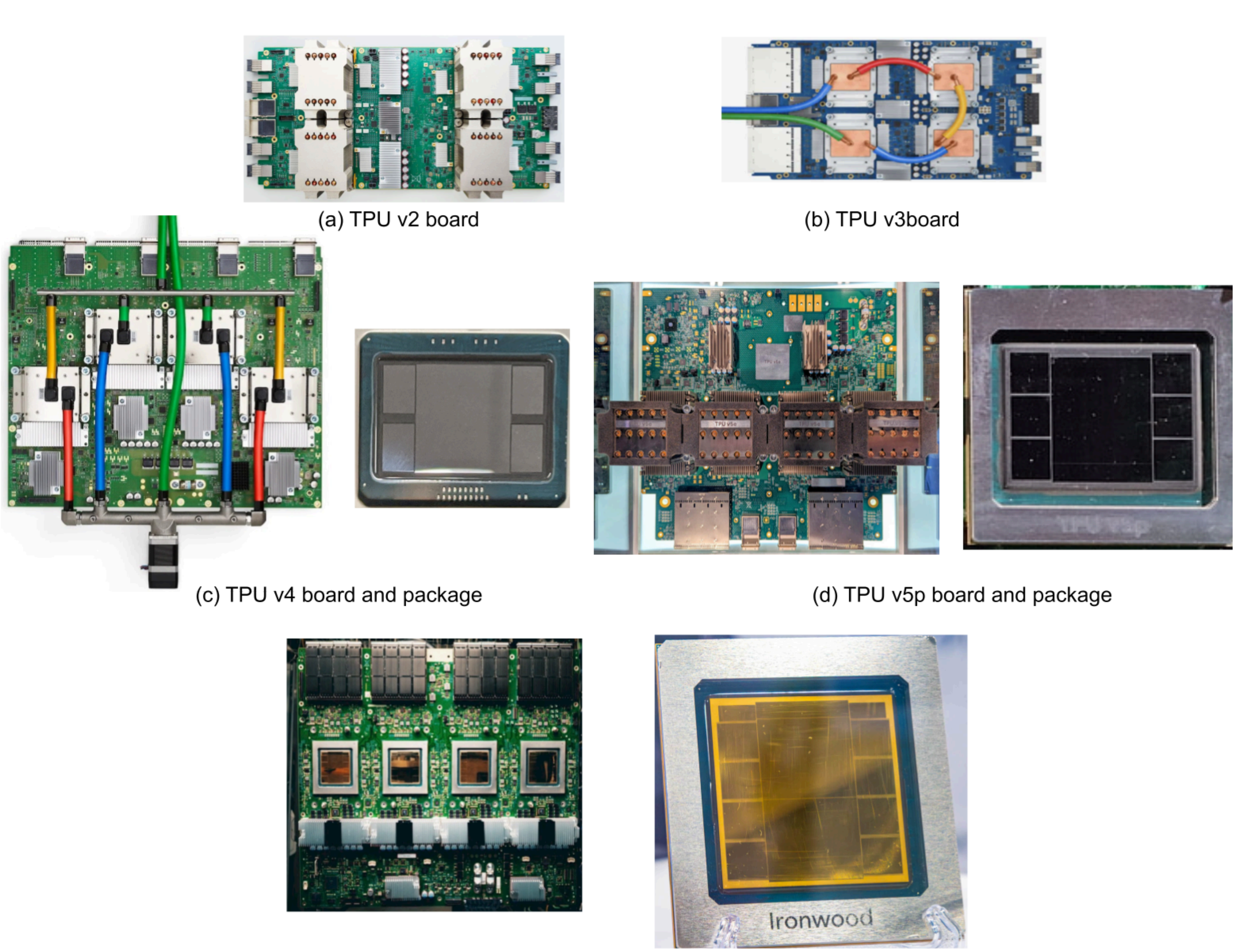

(a) TPU v2 board (b) TPU v3board

(c) TPU v4 board and package (d) TPU v5p board and package

(e) Ironwood board and package

**Figure 3. Five generations of training TPU trays and closeup photos of the three latest generations of packages.** The HBM stacks are visible on the left and right of the compute die(s) in the TPU packages: 4 for TPU v4, 6 for TPU v5p, and 8 for Ironwood. Ironwood uses two compute dies per package.

- **VPU**: Improved logic density enabled the VPU to evolve from two restricted ALUs per vector lane —each capable of some, but not all, ALU operations—to four general-purpose ALUs per lane. Vector registers also doubled in both dimensions, from 8x128 in TPU v2–v5p to 16x256 in Ironwood. The vector architecture is well-suited for non-matrix DNN operations, such as activation functions, softmax, and batch normalization, and quantization functions for lower precision numerics.

- **VLIW instructions:** Specifying instruction level parallelism via VLIW works well. We simply widen the VLIW instruction each generation as TPUs gain more parallel hardware to control, such as more MXUs; Ironwood instructions are >50% wider than TPU v2. DNN programs are not prohibitively large, making extra instruction memory inexpensive, and backwards binary compatibility—required for CPUs—is superfluous.
- **SparseCores**: Their architecture became more general purpose and performance increased per generation—for example, by 2.4x from TPU v5p to Ironwood—plus their number doubled from two in TPU v2 to four in TPU v4, same as Ironwood.
- **VMEM**: SRAM density grew more slowly than logic density. Consequently, size only quadrupled from 32 MB per node in TPU v2 to 128 MB in TPU v5p and Ironwood despite a considerably larger die area.
- **HBM**: Similar to systolic arrays for MXUs, HBM proved a judicious choice for main memory. Standard DRAM was a bottleneck in TPU v1 [Jouppi17], so TPU v2 upped memory bandwidth by 30x with HBM. Over eight years, TPUs scaled capacity and bandwidth another tenfold from 16 GiB at 700 GB/sec using 4 stacks of HBM2 in TPU v2 to 192 GiB at 7300 GB/sec using 8 stacks of HBM3E in Ironwood. Once again, most accelerators either used HBM initially or eventually adopted it.
- **ICI**: Grew from 4 external links at 62 GB/sec[4] per node, forming a 2D Torus interconnect in TPU v2, to 6 external links at 100 GB/sec per node, allowing a 3D Torus in TPU v4, TPU v5p, and Ironwood. Ironwood increased the number of nodes 36X and bisection bandwidth 39X over TPU v2.
- **Supercomputer size**: Grew from 256 nodes in TPU v2 to 9216 nodes in Ironwood (0.25K to 9K).

## Improved Resilience Over Time

Google has long built dependable computing services using distributed systems of commodity components. Our approach to dependable training supercomputers follows those found in high performance computing, as both involve running large, long-duration batch jobs rather than interactive services. The key features are:

1. **Enhanced node quality**: to reduce the chances of failure during execution.
2. **Error checking**: to discover errors during execution.
3. **Checkpoint/restore**: to recover continued execution of long running batch jobs after a failure occurs.
4. **Strict deterministic repeatability requirements**: to aid in system testing and failure detection.
5. **Modular isolation**: To allow the supercomputer to stay in service even if some nodes have failed.

Google employed synchronous data-parallel training to parallelise over multiple 8960-chip TPU v5p pods in multiple data centers for Gemini 2.5 with a goodput of 93% [Gemini25]. A similar measure at a smaller scale for Gemini 1.0 on TPU v4 was 97% [Gemini23]. *Goodput* is short for "good throughput", which in training systems is the rate of good or effective training progress. For example, we might report a training throughput of X for a system in normal operation, but if the system spends 10% of its total time recovering from errors or failures, then the goodput would be 0.9X.

Point 5 significantly improved starting with TPU v4, the first supercomputer to use *optical circuit switches* (*OCSes*) [Jouppi 23]. (The next few paragraphs are derived from that paper.) To enhance data center networking, Google advanced the state-of-the-art in reliability and cost of optical transceivers based on 3D Micro-Electro-Mechanical Systems (MEMS) mirrors that switch in milliseconds. Within a rack, electrical connections offered the best cost-performance. The question then is what size electrically-cabled building block should be used? Given the 3D torus, 3D cubes have the best bisection bandwidth, suggesting 4×4×4 (64 chips) or 8×8×8 (512). With 4 TPU per CPU host, 64 TPU chips and their 16 CPU hosts comfortably fit into one rack. Since 512 chips need multiple racks, a 4x4x4 ($4^3$) building block was chosen; we call them *cubes*.

[4] Bandwidth per direction. Some accelerators would report double the bandwidth, arguing that the link can simultaneously send and receive. We follow network conventions, i.e., Ethernets that simultaneously send and receive at 100 Gbits/second are labeled 100 G-bit, not 200 G-bit.

Figure 4 illustrates the links from the 6 “faces” of a $4^3$ cube. There are 16 links per face, totaling 96 optical links per cube that connect to OCSes. To create the wraparound links of a 3D torus, links on opposing sides must connect to the same OCS. Therefore, each cube connects to 6 × 16 ÷ 2 = 48 OCSes. The TPU v4 OCS featured 136×136 (128 ports plus 8 spares for link testing and repairs), so 48 OCSes connect to 48 pairs of cables from 64 cubes (each 64 chips), resulting in the desired total of 4096 TPU v4 chips. Similar to HPC supercomputers, the workload comprises various scale sizes, termed "*slices*,"i.e., 64, 128, …, 2048 chips.

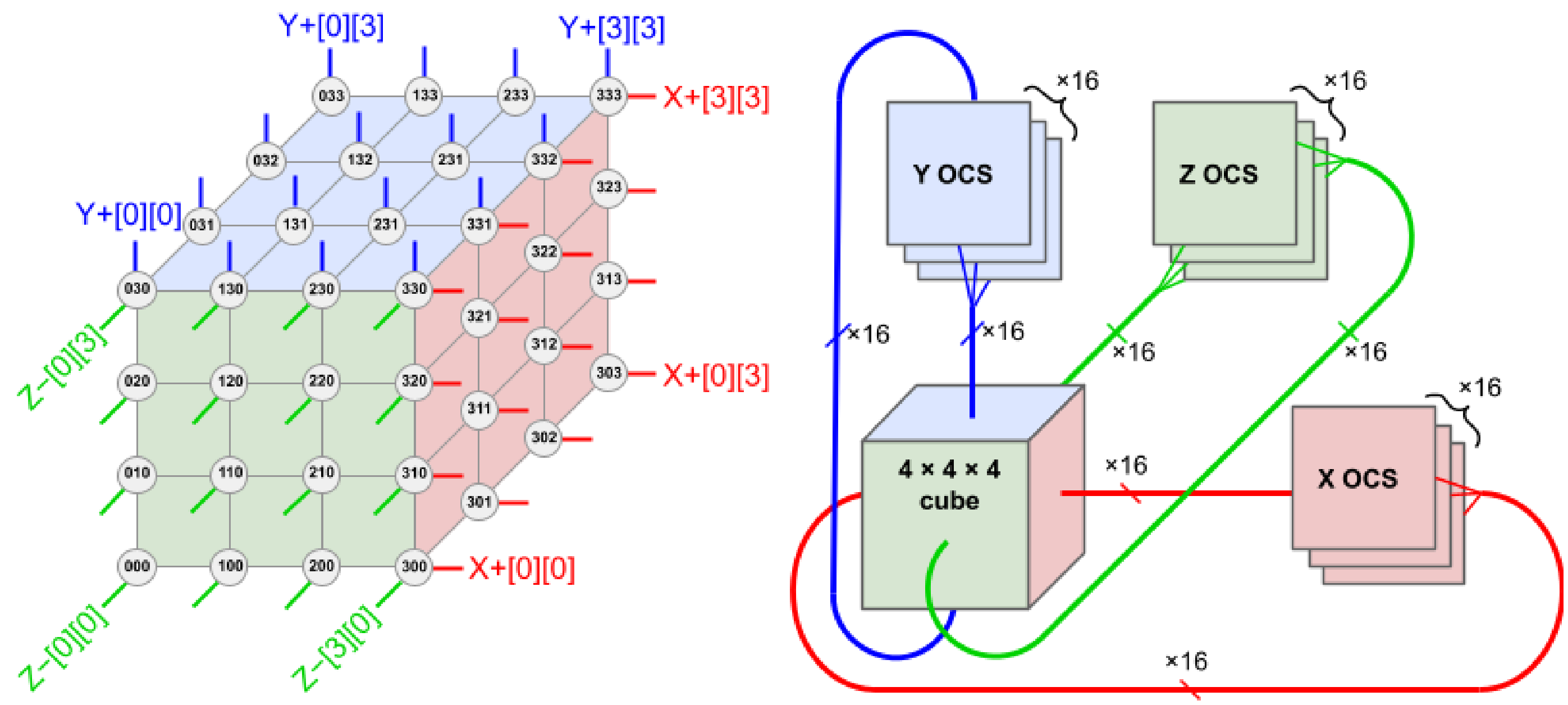

**Figure 4. TPU Cubes and optical interconnect [Jouppi23].**

ICI, the TPU supercomputer interconnect, relies on a 3D Torus topology with a distributed router as part of every TPU chip; it needs no extra chips for TPU-to-TPU communication. OCSes provide essentially a physical layer below ICI that enhances availability by routing around failures. The primary availability challenge for TPU supercomputers lies with the CPU host; each host has 4 TPUs, meaning an Ironwood supercomputer has 2304 CPU hosts. Without OCSes, host availability must be >99.9% to achieve high slice goodput.

The OCS also streamlines scheduling, thereby increasing utilization. For TPU v2 and TPU v3, scheduling a 128 chip slice required the scheduler to locate 128 contiguous idle chips. With OCS, the scheduler can select two $4^3$ cubes from anywhere in the supercomputer. The difficulty of scheduling increases sharply with slice size. This improved availability and simplified scheduling are why Ironwood has 9K nodes rather than a power of 2. Ironwood can run four of the popular 2K slice jobs (each requiring 32 cubes) even if some nodes are down, as 16 spare cubes remain available as substitutes.

OCSes also shrink deployment time [Jouppi23]. TPU v3 systems were not usable until all 1024 chips and all cables were installed and tested. Delivery delays for any component held up the entire supercomputer. Starting with TPU v4, OCSes made each rack independent, so each cube was put into production as soon as 64 chips and the necessary cables were installed and tested. Incremental deployment greatly improves the time to production use and thus cost effectiveness of TPU supercomputers.

*Silent Data Corruption* (*SDC*) in compute logic presents a critical challenge to large-scale AI reliability [George26]. Marginal defects, stemming from manufacturing escapes or silicon aging, can silently degrade model quality and convergence. While previous TPU generations relied on software-based health checks and in-situ workload monitors [Gemini25], Ironwood extends these mitigations into hardware:

1. The **Functional Built-In Self-Test (FBIST)** engine, integrated within the MXU, executes high-coverage functional test patterns during manufacturing and data center burn-in to intercept chips that pass structural tests before they reach the fleet and during operation to find newly flawed chips.
2. To combat intermittent errors in the compute datapath caused by environmental stressors—such as voltage fluctuations, temperature shifts, or specific data patterns—Ironwood introduces a **hardware replay unit** for the VPU. This compiler-transparent unit randomly samples vector bundles for opportunistic redundant execution within existing idle slots in VLIW instructions. By replaying odd-lane operations on the even lanes without altering architectural state, the mechanism provides effective error detection with zero performance overhead and negligible power impact. Deployed across the production fleet, this in-situ monitoring has consistently identified defective units that evaded all other screening methods. Once identified, these units are immediately removed from production using the OCSes and later repaired.

## Improved Power Efficiency Over Time

Vahdat et al. [Vahdat24] recommended that accelerator designers consider average power per workload goodput rather than just benchmark performance per *total cost of ownership* (*TCO*). This shift is driven by the increasing difficulty of securing sufficient electrical power for new data centers, encouraging maximum utilization of available power. Today, *performance per Watt* is more highly regarded than performance per TCO. Figure 5 illustrates a consistent improvement in performance per Watt across TPU generations, although it uses peak performance per TDP (Thermal Dissipation Power) Watt rather than measured performance and power running production workloads as Vahdat et al. recommend. The significance of operational emissions in the overall carbon footprint of TPUs, as we see next, highlights the importance of these gains.

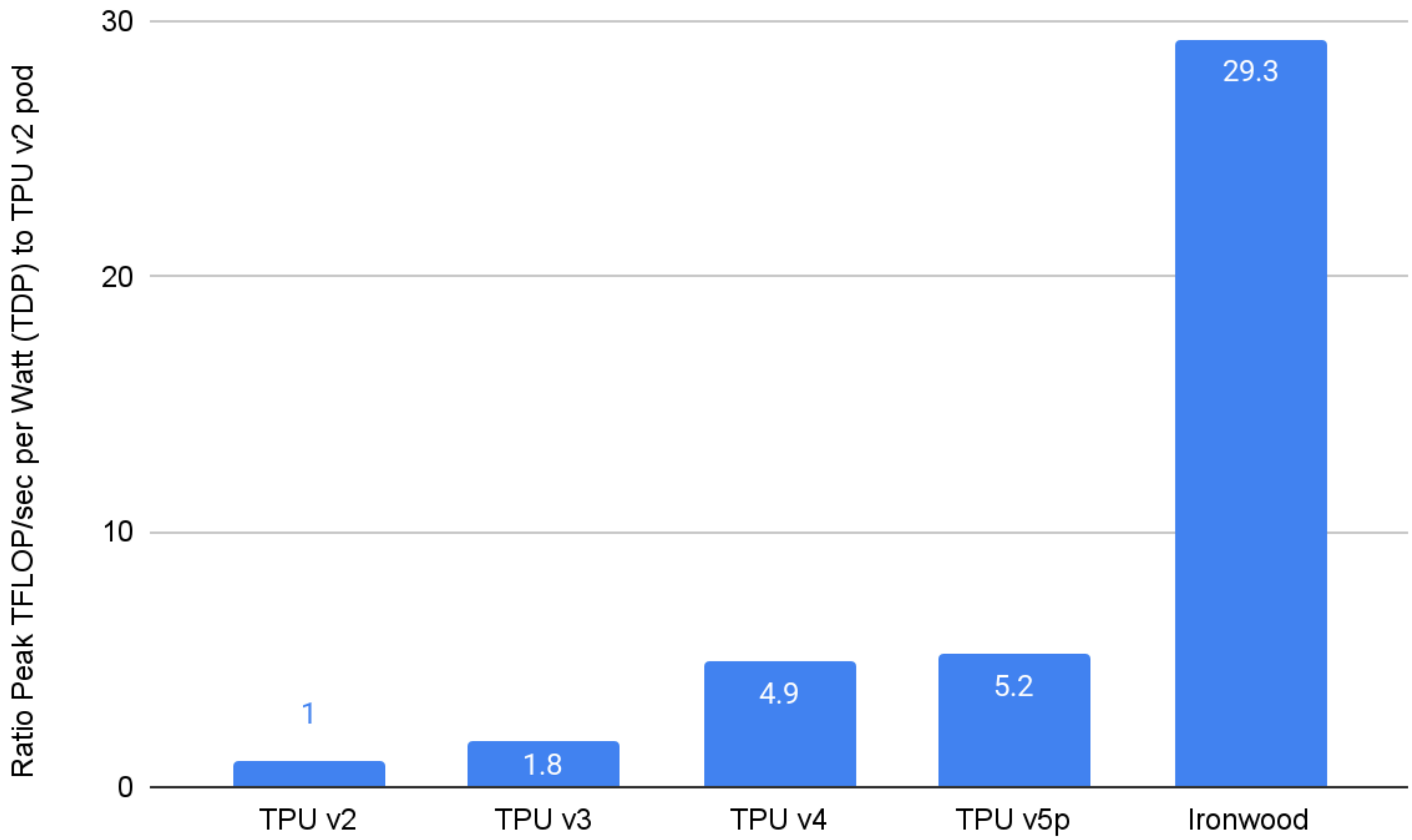

**Figure 5. Relative improvement in peak performance per TDP Watt of TPU supercomputers for training over five generations.** (Bigger is better.)

## Improved Sustainability Over Time

Vahdat et al. also recommended that attention be paid to carbon emissions per workload goodput [Vahdat 24]. Toward that end, Google recently completed a *life-cycle assessment* (*LCA*) of several TPUs [Schneider25]. LCA is a comprehensive analysis of the greenhouse gas (GHG) emissions associated with TPU hardware that includes the entire lifespan of the hardware, from the extraction of raw materials to manufacturing and energy use. (This section is derived from that paper.)

A key challenge was balancing the increased manufacturing costs and power consumption of newer TPUs with their growing performance. Newer TPUs might consume more power per second, but training also takes fewer seconds. The answer was a new metric: *compute carbon intensity* (*CCI*). It quantifies the *carbon dioxide equivalent emissions*[5] (*$CO_2e$*) per utilized floating point operation, or $CO_2e$/FLOP[6]. CCI's significant advantage is that iit's easy to incorporate both embodied carbon emissions and operational carbon emissions, as total CCI = operational CCI + embodied CCI. Unlike CCI, performance per Watt excludes embodied emissions.

Figure 6 shows the CCI for three training TPUs [Schneider25][Huang26]. TPU v5p is 1.1x better for operational and overall CCI than TPU v4, and 1.3x for embodied CCI. Ironwood has a much bigger operational jump of ~3.7x, as Figure 5 suggests, and ~3.8x for embodied.

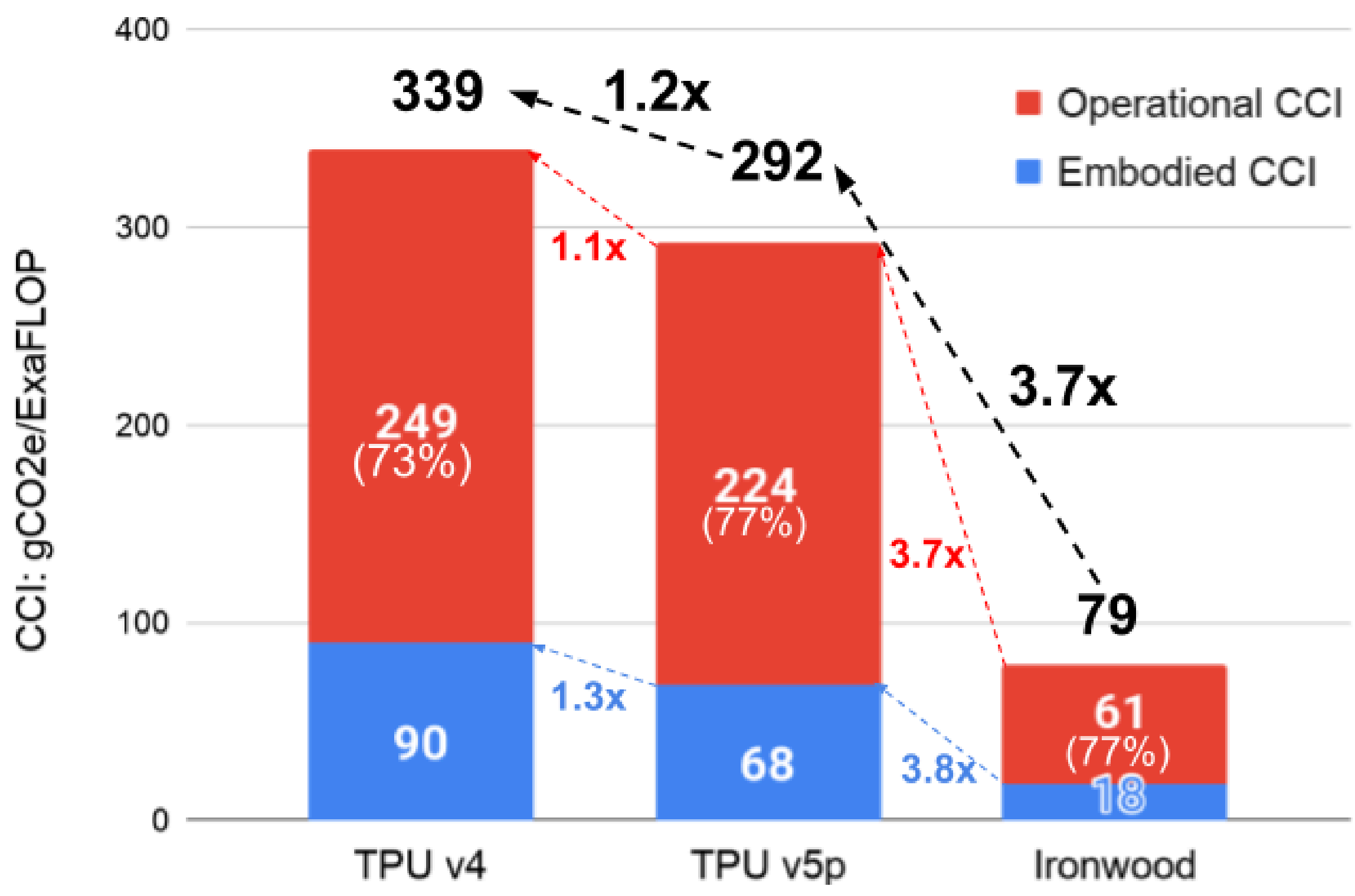

**Figure 6. Compute carbon intensity[5] ($gCO_2e/10^{18}$ FLOPs) for three training TPUs.[7]** (Smaller is better.) Embodied emissions are amortized over six year lifetimes for all TPUs. While the absolute embodied CO2e is higher for Ironwood vs TPU v5p and higher for TPU v5p vs TPU v4, the carbon emissions *per floating point operation* (CCI) is lower because the newer TPUs are so much faster than their predecessors.

---

[5] *Carbon dioxide equivalent* (*$CO_2e$*) measures the climate impact of GHGes like methane and nitrous oxide by converting them to the amount of *$CO_2$* that would cause the same amount of warming over, say, 100 years, using their global warming potential. Combining GHGes into one number simplifies carbon assessment.

[6] ExaFLOP ($10^{18}$ FLOPs) was picked so that $CO_2e$ is in grams versus a smaller unit. It is a fixed amount of computation, not a rate: it's ExaFLOP, not ExaFLOPs/second (aka ExaFLOPS).

[7] The GHG protocol offers two variations of operational emissions. Figure 6 shows *market-based emissions*, which credits carbon-free energy purchases. *Location-based*, which excludes them, would raise operational CCI to 793, 712, and 195 gCO2e/EFLOP, respectively. CCI ratios between TPUs don't change, but Ironwood's embodied CCI would drop from ~23% to ~8% of total emissions.

Operational CCI is ~75% of overall CCI for all three TPUs, reflecting both the relatively high power use and long lifetimes of data center AI accelerators. Unsurprisingly, given the low power and short lifetimes of mobile devices, the roles reverse: 87% of smartphone emissions are embodied [Patterson24].

By the way, operational CCI subsumes performance per Watt since

$$Operational\ CCI\ = \frac{Electricity\ Emissions\ Factor}{Performance/\ Watt\ (measured)}$$

*Electricity Emissions Factor* represents the amount of GHG emitted per unit of electricity consumed (e.g., $gCO_2e/kWh$).

Given FLOPs for a task, CCI also offers a ballpark estimate of emissions. For example, training GPT-3 took ~$3.14 \times 10^{23}$ FLOPs and TPU v5p's CCI is $265 \times 10^{-18}$ $gCO_2e$, so emissions is just their product: ~$83 \times 10^{6}$ $gCO_2e$ or ~83 million metric tons of carbon dioxide equivalent emissions ($mtCO_2e$).

Performance per Watt only helps with operational emissions while CCI specifies performance per Watt *and* embodied emissions *and* makes it easy to estimate carbon emissions.

## Conclusion: Architectural Stability For Five Generations Despite Improved Scale, Resilience, Power Efficiency, and Sustainability

This paper recaps the evolution and lasting effectiveness of Google's training TPUs across five generations, from TPU v2 to Ironwood (TPU 7). Key advancements over an eight-year period demonstrate significant scaling despite the end of Dennard Scaling and the decline of Moore's Law: HBM capacity and bandwidth increased 10x, supercomputer size grew 36X, and peak TPU node performance improved 100x, which multiplied together yields a 3600x increase in peak supercomputer performance. These gains were achieved while also enhancing resilience.

Optical circuit switches (OCSes) significantly reduce deployment time and improve scheduling efficiency by enabling modular installation and fault isolation. Another benefit is to improve availability by mapping out failures to restore the 3D Torus topology. Without OCS, we would likely need a failure strategy per rack, likely necessitating a very different network topology. The downsides of this alternative include increased TCO [Jouppi23] and exacerbating the "rackification" and subsequent cooling problems by making each rack include spare TPUs and standalone crossbar switches, which could use nearly half the rack space. OCSes also allow any number of spare cubes for resilience, which can be adjusted based on component reliability in the field vs when deployed with a fixed number of spare TPUs per rack. Most importantly, OCSes simplify programming by maintaining a single tier network across the pod while allowing the swapping out of failed components instead of having to program a two-tier network with different bandwidth and latency per level.

Ironwood also introduced hardware-based mitigations—functional built-in self-test and hardware replay—to combat silent data corruption and to ensure higher reliability in compute datapath operations.

In this era of limited electricity for new data centers, performance per Watt superseded performance per TCO. The 30x peak performance per TDP Watt improvement across generations—notably 6X for Ironwood from TPU v5p—indicates Google's ongoing dedication to reduced environmental impact alongside increased computational power. We encourage architects to embrace *compute carbon intensity* (*CCI*). CCI reflects a more holistic approach—as it encompasses both embodied and operational emissions whereas performance per Watt only covers operational emissions—and it simplifies estimates of environmental impact of AI.

Despite initial skepticism regarding the longevity of an ASIC domain-specific architecture given rapidly evolving DNNs, the initial TPU v2 microarchitecture demonstrated remarkable stability and adaptability. Stability means model, compiler, and software stack optimizations from prior TPUs are easily reused for new TPUs. TPUs have successfully accommodated rapidly changing and diverse workloads over a decade, including Transformers and Diffusion. Listed by order of importance, six key initial decisions have proven their worth through their longevity in TPUs and by their adoption by other AI training accelerators that followed:

1. **Systolic Arrays** for matrix multiplies.
2. **Range-oriented, narrow floating point arithmetic** (BF16, FP8, FP4) versus precision-oriented, wide IEEE floating point arithmetic (FP16, FP32, FP64).
3. **HBM** for main memory.
4. **Custom high speed links** (in our case **ICI**) to assemble AI accelerators into AI supercomputers.
5. **Software control of the memory hierarchy via DMAs** with scratchpad SRAM versus a conventional CPU-like cache hierarchy.
6. **Vector Units** to perform the non matrix computations.

To our best knowledge, two other innovations—OCSes and SparseCores—remain unique to TPUs. The TPU's VLIW instruction word allowed our compilers and software ecosystem to deliver optimized compilation performance with deterministic timing; while some accelerators also use VLIW, others picked very different approaches to control. Google has used liquid cooling exclusively for training since TPU v3 in 2018, while others generally come in both liquid-cooled and slower, lower-power, air-cooled versions. And there is no agreement on the number of processors per accelerator; a conservative range is at least 2 to 1472 [Jouppi23].

In the 1960s, conventional CPU wisdom converged on microcoded, backwards binary compatible instruction sets with cache memory—which the IBM 360 pioneered—then switched in the 1990s to superscalar, speculative, out-of-order architectures after the pathbreaking Intel Pentium Pro. Perhaps the six features above will be considered the defining features of training accelerators of the 2020s? We hope that the TPU architecture, designed for AI, will follow the IBM 360 and x86 that many decades later are still useful for their intended markets, business data processing and personal computing.

In conclusion, TPUs offer a potent combination of five strengths:

1. Having **only two large processors per TPU**—which compilers can unify to offer an even larger single Megacore [JAX24]—that gobble huge chunks of data **simplifies developing new models** versus programming numerous tiny processors that nibble small bites of data. As supercomputer legend Seymour Cray famously asked: "If you were plowing a field, which would you rather use: two strong oxen or 1024 chickens?"
2. **Architectural stability plus hardware/software codesign *inside a single organization* means tuned models are available immediately as new TPUs debut** versus a much larger external effort that can't start until new accelerators are officially released.
3. **O(10,000) supercomputer size in a single tier network with enhanced resilience allows training of long-running giant models across multiple supercomputers in multiple regions at >90% goodput** simply by using data parallelism [Gemini25].
4. A track record of **increasing compute, memory capacity, memory and interconnect bandwidth, and supercomputer size each generation** for five generations.
5. **Reduced environmental impact** via decreased emissions per floating point operation [Schneider25] [Huang26].

These advantages position TPUs to deliver the computational requirements needed to sustainably unleash AI's huge potential.

# Acknowledgments

We extend our sincere gratitude to the numerous Googlers who developed TPU hardware and software over the past decade. Thanks to Ramdas Nagarajan for help on SDC and FBIST, to Ravi Ragnar for the latest DNN mix, and for reviews from Prashant Chandra, James Laudon, Parthasarathy Ranganathan, and the anonymous *Micro* reviewers.